\documentclass[conference]{IEEEtran}
\IEEEoverridecommandlockouts

\usepackage{cite}

\usepackage{amsmath,amssymb,amsfonts}
\usepackage{mathtools}
\usepackage{amsthm}

\usepackage{graphicx}
\usepackage{xcolor}
\usepackage{booktabs}
\usepackage{longtable}
\usepackage{array}
\usepackage{makecell}
\usepackage{pdflscape}
\usepackage{pgfplots}
\pgfplotsset{compat=1.18}

\usepackage{algorithm}
\usepackage{algpseudocode}

\usepackage{listings}
\usepackage{pifont}
\usepackage{textcomp}
\usepackage{multicol}
\usepackage{ifthen}
\usepackage{soul}
\usepackage{comment}

\usepackage{balance}
\usepackage[colorlinks=true, citecolor=blue]{hyperref}
\def\BibTeX{{\rm B\kern-.05em{\sc i\kern-.025em b}\kern-.08em
    T\kern-.1667em\lower.7ex\hbox{E}\kern-.125emX}}
\begin{document}

\title{ Agentic AI Cybersecurity Framework\\
}
\author{ 
\IEEEauthorblockN{Victor Kebande\IEEEauthorrefmark{1}\IEEEauthorrefmark{2}}
\IEEEauthorblockA{\IEEEauthorrefmark{1}University of Colorado Denver, CO USA\\
\IEEEauthorrefmark{2}ATLAS Institute, University of Colorado 
Boulder, Colorado, USA\\
Email: victor.kebande@ucdenver.edu, victor.kebande@colorado.edu}
} 

\maketitle

\begin{abstract}
The increasing scale, complexity, and dynamism of modern cyber threats have rendered traditional reactive cybersecurity mechanisms insufficient. This paper introduces an \textit{Agentic AI Cybersecurity Framework (AACF)} designed to enable autonomous, goal-driven, and adaptive cyber defense operations. Unlike conventional systems that rely on predefined rules and human intervention, the proposed framework leverages agentic artificial intelligence to perceive environmental states, reason about potential threats, and execute context-aware responses with minimal supervision. The framework is structured into key functional layers, including perception, reasoning, decision-making, action execution, and feedback-driven learning, enabling continuous adaptation to evolving attack patterns. By integrating intelligent agents with real-time data analysis and automated response mechanisms, AACF supports proactive threat detection, dynamic risk assessment, and coordinated mitigation strategies across distributed environments. A conceptual architecture is presented, along with illustrative use cases demonstrating its applicability to intrusion detection, incident response, and autonomous security orchestration. The proposed framework contributes to the emerging paradigm of self-directed cybersecurity systems and provides a foundation for developing resilient, scalable, and intelligent defense infrastructures.
\end{abstract}


\section{Introduction}

The rapid and continous evolution of cyber threats, coupled with the increasing complexity of modern digital infrastructures, has significantly challenged traditional cybersecurity approaches. Conventional security mechanisms are largely reactive, relying on predefined rules, signature-based detection, and human intervention to identify and mitigate attacks \cite{edge2009survey}. While effective against known threats, these approaches struggle to cope with sophisticated, adaptive, and large-scale attacks such as advanced persistent threats (APTs), zero-day exploits, and AI-driven cyber intrusions \cite{khule2025layered}. As a result, there is a growing need for intelligent, autonomous, and adaptive cybersecurity systems capable of responding to dynamic threat environments in real time.

Recent advances in Artificial Intelligence (AI) and Machine Learning (ML) have introduced new possibilities for enhancing cybersecurity capabilities. AI-driven systems have demonstrated success in tasks such as anomaly detection, intrusion detection, malware classification, and threat intelligence analysis. However, most existing AI-based cybersecurity solutions remain fundamentally reactive and narrowly scoped, often limited to specific tasks without the ability to autonomously plan, reason, and execute complex defense strategies. These limitations highlight the need for a paradigm shift from reactive intelligence to proactive, goal-driven autonomy in cybersecurity systems.

Agentic AI \textit{(aAI)} has recently emerged as a promising paradigm that extends beyond traditional AI by enabling systems to operate as autonomous agents capable of perceiving their environment, reasoning about objectives, making decisions, and executing actions with minimal human intervention \cite{acharya2025agentic}. Unlike conventional AI models that respond to isolated inputs, agentic systems are goal-oriented, iterative, and capable of adapting their behavior based on feedback and environmental changes. This capability makes \textit{aAI} particularly well-suited for cybersecurity applications, where threats are dynamic, adversarial, and continuously evolving.

From a cybersecurity perspective, the integration of agentic AI introduces the possibility of developing self-directed defense systems that can autonomously detect, analyze, and respond to threats across distributed environments. Such systems can continuously monitor network activities, correlate multi-source data, assess risks, and orchestrate mitigation strategies in real time. By leveraging autonomous decision-making and adaptive learning, agentic AI can enhance the resilience, scalability, and responsiveness of cybersecurity infrastructures.

In this paper, we propose a generic  \textit{Agentic AI Cybersecurity Framework (AACF)} that enables autonomous and intelligent cyber defense operations. The proposed \textit{AACF} framework is designed around key functional layers; perception, reasoning, decision-making, action execution, and feedback-driven learning. These layers collectively support continuous situational awareness, dynamic threat analysis, and coordinated response mechanisms. The \textit{AACF} framework aims to bridge the gap between traditional AI-based security systems and fully autonomous cybersecurity solutions by introducing goal-driven intelligence and adaptive behavior into cyber defense processes.

The main contributions of this work can be summarized as follows:
\begin{itemize}
    \item We introduce a novel Agentic AI Cybersecurity Framework \textit{AACF} for autonomous and adaptive cyber defense.
    \item We define a \textit{AACF}-layered architecture that integrates perception, reasoning, decision-making, action, and feedback mechanisms.
    \item We demonstrate how agentic AI enables proactive threat detection, dynamic risk assessment, and automated response.
    \item We provide illustrative use cases highlighting the applicability of the proposed framework in real-world cybersecurity scenarios.
\end{itemize}

The remainder of this paper is organized as follows. Section II and III presents background and related work respectively.  Section IV describes the proposed Agentic AI Cybersecurity Framework in detail. Section V presents a use case and application scenarios. Section VI discusses implications, limitations, and future research directions, and Section VII concludes the paper.

\section{Background }

\subsection{ Agentic AI}

Agentic AI represents a significant evolution in artificial intelligence, shifting from reactive, prompt-driven systems to autonomous, goal-oriented entities capable of reasoning, planning, and acting with minimal human intervention \cite{xu2025integrating, rusell2010artificial}. Unlike traditional AI models that operate within predefined constraints, agentic systems exhibit autonomy, adaptability, and decision-making capabilities, enabling them to execute complex, multi-step tasks in dynamic environments \cite{xu2025integrating}. These systems typically operate in iterative cycles involving perception, reasoning, action, and learning, allowing continuous adaptation based on environmental feedback.

From a systems perspective, agentic AI often leverages large language models (LLMs) integrated with external tools, memory modules, and orchestration mechanisms to enable long-horizon task execution and coordination across multiple agents~\cite{wei2026agentic}. This architectural shift enables AI systems to move beyond isolated predictions toward sustained workflows, making them suitable for domains such as cybersecurity, where real-time decision-making and adaptability are critical.

\subsection{AI in Cybersecurity}

Artificial intelligence has been widely adopted in cybersecurity for tasks such as intrusion detection, anomaly detection, malware classification, and threat intelligence analysis \cite{becue2021artificial}. Machine learning models have demonstrated effectiveness in identifying patterns and anomalies in large-scale datasets, improving detection accuracy and reducing response time. However, most existing AI-driven cybersecurity systems are limited to task-specific functions and rely heavily on human analysts for interpretation and action.

Recent developments have highlighted the limitations of these approaches, particularly in dealing with advanced and rapidly evolving threats such as zero-day exploits and advanced persistent threats (APTs). Traditional systems often lack the ability to autonomously correlate multi-source data, reason about attack contexts, and orchestrate coordinated responses. As a result, there is increasing interest in transitioning from reactive AI systems to proactive and autonomous cybersecurity solutions.

\subsection{Agentic AI in Cybersecurity}

The integration of agentic AI into cybersecurity has emerged as a promising direction for enabling autonomous cyber defense. Agentic AI systems can continuously monitor environments, proactively identify vulnerabilities, and orchestrate responses to security incidents without requiring constant human supervision~\cite{evani5332681agentic}. In security operations centers (SOCs), such systems can automate alert triage, accelerate investigations, and coordinate responses across multiple tools and platforms.

Recent studies have explored the use of multi-agent systems and autonomous pipelines for cybersecurity operations. These systems combine reasoning, tool usage, and memory to perform complex investigative workflows, moving beyond traditional detection pipelines~\cite{vinay2026evolution}. Furthermore, agentic AI enables capabilities such as autonomous threat hunting, dynamic risk assessment, and continuous security validation, significantly enhancing operational efficiency and scalability \cite{leo2026threat}.

However, this rapid adoption of agentic AI also introduces new challenges. The autonomy and adaptability of these systems create novel attack surfaces, including manipulation of reasoning processes, memory poisoning, and exploitation of tool integrations. Additionally, the dual-use nature of agentic AI enables adversaries to leverage similar capabilities for automated reconnaissance, exploitation, and large-scale cyberattacks.

\section{Related Work}

Recent research has begun to formalize the role of agentic AI in cybersecurity. Several studies propose conceptual frameworks and architectures that model cybersecurity systems as distributed, multi-agent environments capable of autonomous decision-making and coordination. For instance, meta-cognitive architectures have been introduced to govern agent behavior and ensure accountability in autonomous cyber defense systems~\cite{tkach2025towards}. These approaches emphasize the importance of integrating reasoning, explanation, and governance mechanisms to manage uncertainty and risk.

Survey studies have further highlighted the rapid evolution of agentic AI in cybersecurity, from single-model systems to complex multi-agent ecosystems and autonomous pipelines~\cite{vinay2026evolution}. These works identify key research challenges, including coordination among agents, validation of autonomous actions, explainability, and the need for robust safety and governance frameworks. Other research by \cite{kebande2026end} has projected the end of traditional reasoning (Pre-training) to a more agentic inspired reasoning, which shows the need for cybersecurity in superintelligent systems.

\begin{figure*}[h!]
\centering
\includegraphics[width=0.9\textwidth]{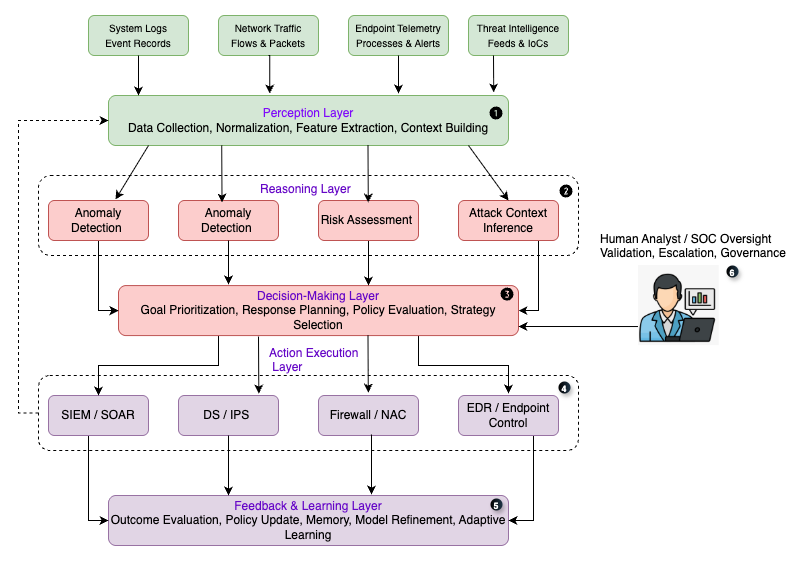}
\caption{Proposed Agentic AI Cybersecurity Framework (AACF)}
\label{fig:myimage}
\end{figure*}

Despite these advancements, existing approaches remain fragmented and often focus on specific components such as detection, response, or governance. There is a lack of unified frameworks that integrate perception, reasoning, decision-making, action, and learning into a cohesive architecture for autonomous cybersecurity. This gap motivates the need for a comprehensive Agentic AI Cybersecurity Framework that systematically addresses these dimensions and enables scalable, adaptive, and intelligent cyber defense systems.

\section{Proposed Agentic AI Cybersecurity Framework}

This section presents the proposed \textit{Agentic AI Cybersecurity Framework (AACF)}, a unified architecture designed to enable autonomous, adaptive, and goal-driven cybersecurity operations. The  \textit{AACF} framework integrates perception, reasoning, decision-making, action execution, and feedback-driven learning into a continuous closed-loop system capable of responding to dynamic and adversarial environments. Figure~\ref{fig:myimage} labeled 1-6 illustrates the architecture of the proposed \textit{AACF}. The \textit{AACF} framework operates as a closed-loop system in which data flows from perception to reasoning, decision-making, and action execution, while continuously being refined through feedback and learning.

\subsection{Perception Layer (1)}

The perception layer labeled 1 serves as the entry point of the \textit{AACF} framework and is responsible for acquiring and preprocessing heterogeneous cybersecurity data. As shown in Figure~\ref{fig:myimage}, this layer integrates multiple data sources, including system logs, network traffic flows, endpoint telemetry, and threat intelligence feeds. These inputs capture both internal system activities and external threat indicators, enabling comprehensive visibility of the operational environment.

In the perception layer, raw data undergoes transformation through data collection, normalization, feature extraction, and context-building processes. Data normalization ensures consistency across heterogeneous sources, while feature extraction identifies relevant attributes for downstream analysis. Context building aggregates these features into structured representations that reflect the state of the system and the behavioral patterns. This layer establishes the foundation for situational awareness by providing a coherent and unified view of the cybersecurity environment.

\subsection{Reasoning Layer (2)}

The reasoning layer labeled 2 performs a contextual analysis of the processed data to identify threats and infer adversarial behavior. As depicted in Figure~\ref{fig:myimage}, this layer consists of four tightly integrated components that collectively enable deep understanding of security events.

The anomaly detection component in Figure~\ref{fig:myimage} identifies deviations from normal system behavior using statistical and machine learning techniques, enabling early detection of potential threats. The event correlation component aggregates and links related events across multiple data sources, transforming isolated observations into coherent attack sequences. The risk assessment component evaluates the severity and potential impact of detected threats by considering factors such as asset criticality, attack likelihood, and potential damage. The attack context inference component constructs a higher-level understanding of adversarial intent, tactics, and progression paths, allowing the system to anticipate future actions.

These components enable the framework to transition from reactive detection to proactive and context-aware threat analysis.

\subsection{Decision-Making Layer (3)}

The decision-making layer labeled 3 is responsible for determining appropriate response strategies based on insights generated by the reasoning layer. As shown in Figure~\ref{fig:myimage}, this layer performs goal prioritization, response planning, policy evaluation, and strategy selection.

Goal prioritization ensures that critical threats are addressed first, while response planning evaluates multiple mitigation options in relation to operational constraints. Policy evaluation ensures that selected actions comply with organizational policies and security guidelines. Strategy selection identifies the most effective course of action to mitigate threats while minimizing disruption to system operations.

This layer embodies the agentic intelligence of the framework, enabling dynamic and adaptive decision-making in response to evolving threat conditions.

\subsection{Action Execution Layer (4)}

The action execution layer labeled 4 implements the decisions generated by the decision-making layer by interfacing with operational cybersecurity tools and infrastructure. As illustrated in Figure~\ref{fig:myimage}, this layer integrates with SIEM and SOAR platforms, intrusion detection and prevention systems, firewalls and network access control mechanisms, and endpoint detection and response systems.

Based on these integrations, the \textit{AACF}  framework can autonomously execute mitigation actions such as blocking malicious traffic, isolating compromised endpoints, updating security policies, and initiating incident response workflows. The ability to perform these actions in real time significantly reduces response latency and enhances the effectiveness of cybersecurity operations.

\subsection{Feedback and Learning Layer (5)}

The feedback and learning layer labeled 5 in \textit{AACF} framework is responsible for evaluating the outcomes of executed actions and refining the system’s behavior accordingly. As is shown in Figure~\ref{fig:myimage}, this layer performs outcome evaluation, policy updates, memory management, model refinement, and adaptive learning.

The outcome evaluation assesses the effectiveness of mitigation strategies, while policy updates adjust decision-making rules based on observed results. Memory mechanisms store historical information, enabling the system to learn from past experiences. Model refinement continuously improves detection and reasoning capabilities through retraining and adaptation. The feedback loop connects this layer back to the perception layer, enabling continuous learning and system evolution.

\subsection{Human Analyst Oversight (6)}

The \textit{AACF} framework incorporates human analyst oversight to ensure accountability, governance, and validation of critical decisions. As shown in Figure~\ref{fig:myimage}, human analysts in Step 6 interact with the decision-making and feedback layers, enabling escalation and review of complex or high-risk scenarios. This integration ensures that the system maintains a balance between autonomy and human control, allowing expert intervention when necessary while preserving the efficiency of automated operations.

\subsection{Operational Flow}

The AACF operates as a continuous closed-loop system in which perception informs reasoning, reasoning guides decision-making, decisions trigger actions, and outcomes are evaluated through feedback and learning. This iterative process enables continuous adaptation and supports proactive cybersecurity operations in dynamic environments.

\section{Use Case: Autonomous Intrusion Detection and Response}

To demonstrate the applicability of the proposed \textit{AACF} framework, a representative use case involving autonomous intrusion detection and response within an enterprise environment is considered. In such an environments large volumes of heterogeneous data are continuously generated, often overwhelming traditional security systems and human analysts.

Within the AACF, the perception layer continuously ingests data from network traffic, system logs, endpoint telemetry, and threat intelligence feeds. When anomalous behavior is detected, the reasoning layer correlates these observations and constructs a contextual understanding of the potential threat. For example, abnormal login attempts combined with unusual lateral movement may indicate a compromised system.

The decision-making layer evaluates this context and determines an appropriate response strategy, which is then automatically executed through integrated security tools in the action execution layer. The feedback and learning layer subsequently evaluates the outcome and updates system behavior, while human analysts may intervene when necessary. This demonstrates the framework’s ability to provide rapid, coordinated, and adaptive responses.

\section{Discussion}

The proposed Agentic AI Cybersecurity, \textit{AACF}, Framework introduces a paradigm shift toward autonomous and intelligent cyber defense systems. By integrating multiple layers of intelligence into a unified architecture, the \textit{AACF} framework enhances situational awareness, reduces response latency, and enables continuous adaptation.

However, challenges remain in ensuring robustness against adversarial manipulation, maintaining explainability of autonomous decisions, and achieving seamless integration with existing infrastructure. The inclusion of human oversight in Step 6 of the \textit{AACF} framework provides an important mechanism for governance and validation, although balancing autonomy and control remains an open research challenge.

The primary advantage of the \textit{AACF} framework lies in its ability to enable autonomous cybersecurity operations. By reducing reliance on manual analysis, the framework can significantly decrease response times and improve the scalability of security operations. The integration of multi-source data and contextual reasoning enhances situational awareness, allowing the system to detect complex and multi-stage attacks that may be overlooked by traditional approaches.

Despite its advantages, the adoption of agentic AI in cybersecurity introduces new attack surfaces. Autonomous systems that rely on AI models and external data sources are susceptible to adversarial manipulation \cite{sifakis2023trustworthy}, for example data poisoning, model evasion, and prompt injection attacks \cite{shao2025enhancing}. An attacker may attempt to influence the reasoning process by injecting misleading inputs or manipulating contextual data, potentially leading to incorrect decisions or inappropriate actions.

Another critical challenge is the explainability of agentic AI systems. Autonomous decision-making processes can be difficult to interpret, particularly when complex models and multi-step reasoning are involved. In cybersecurity, where decisions may have significant operational and legal implications, the ability to explain and justify actions is essential.

While the \textit{AACF} framework  emphasizes autonomy, human oversight remains a critical component. Fully autonomous systems may not always account for contextual nuances, organizational policies, or ethical considerations. Incorporating a Human-in-the-Loop (HiL) approach ensures that critical decisions can be validated, and ambiguous situations can be escalated for expert analysis.

\section{Conclusion}

This paper presented a the \textit{AACF} framework, a layered architecture designed to enable autonomous, adaptive, and intelligent cybersecurity operations. By integrating perception, reasoning, decision-making, action execution, and feedback-driven learning, the framework supports proactive threat detection and coordinated response.

The proposed the \textit{AACF} framework provides a foundation for next-generation cybersecurity systems capable of operating in dynamic and adversarial environments. Future work will focus on enhancing robustness, improving explainability, and extending the framework toward multi-agent and decentralized cybersecurity architectures and providing a proof-of-concept.

\section*{Acknowledgment}

The author would like to thank anonymous reviewers for their
valuable insights, and
the Department of Computer Science at University of Colorado Denver, USA for their
support while coming up with this research. The author also
acknowledges that the opinions, findings, and conclusions
expressed in this paper are purely of the author.


\balance

\bibliographystyle{IEEEtran}
\bibliography{name}

\end{document}